\documentclass[twocolumn,epjc3]{svjour3}

\usepackage{amsmath}
\usepackage{euscript,amssymb,epsfig}
\usepackage{graphicx}
\usepackage{amsfonts,latexsym}

\usepackage{cite}

\RequirePackage{graphicx}

\newcommand{\be}[1]{\begin{equation}\label{#1}}
\newcommand{\ee}{\end{equation}}
\newcommand{\ba}[1]{\begin{eqnarray}\label{#1}}
\newcommand{\ea}{\end{eqnarray}}
\newcommand{\rf}[1]{(\ref{#1})}
\newcommand{\nn}{\nonumber}

\journalname{Eur. Phys. J. C}

\begin{document}

\title{Zero average values of cosmological perturbations as an indispensable condition for the theory and simulations}

\author{Maxim Eingorn\thanksref{e1,addr1} \and Maxim Brilenkov\thanksref{e2,addr2} \and Branislav Vlahovic\thanksref{e3,addr1}}

\thankstext{e1}{e-mail: maxim.eingorn@gmail.com}

\thankstext{e2}{e-mail: maxim.brilenkov@gmail.com}

\thankstext{e3}{e-mail: vlahovic@nccu.edu}

\institute{CREST and NASA Research Centers, North Carolina Central University,\\ Fayetteville st. 1801, Durham, North Carolina 27707, U.S.A.\\ \label{addr1}
\and Department of Theoretical Physics and Astronomical Observatory, Odessa National University,\\ Dvoryanskaya st. 2, Odessa 65082, Ukraine \label{addr2}}

\date{Received: date / Accepted: date}

\maketitle

\begin{abstract} We point out a weak side of the commonly used determination of scalar cosmological perturbations lying in the fact that their average values
can be nonzero for some matter distributions. It is shown that introduction of the finite-range gravitational potential instead of the infinite-range one
resolves this problem. The concrete illustrative density profile is investigated in detail in this connection.\end{abstract}





\section{Introduction} \setcounter{equation}{0}

As is generally known, if a certain theory uses an average value $\bar f$ of a physical quantity $f$ as its zero-order approximation ($f\approx \bar f$) and a
deviation from this value $\delta f=f-\bar f$ as a quantity of the first order of smallness, then the average value of this deviation $\overline{\delta f}$ is
equal to zero: $\overline{\delta f}=\overline{\left(f-\bar f\right)}=0$. This clear argumentation is relevant, in particular, in the context of cosmological
perturbations, if we assume that the homogeneous Friedmann-Lemaitre-Robertson-Walker (FLRW) geometry represents an average geometry for our cosmological
spacetime. Indeed, proceeding on this assumption, the scale factor, which enters into the background FLRW metric and does not depend on the spatial location,
actually describes the averaged metric coefficients originating from the averaged material sources. Here and in what follows we imply averaging over the volume in
the comoving coordinates. Therefore, one immediately arrives at the inevitable conclusion that when the first-order perturbation theory is constructed against the
homogeneous FLRW background, the spatial averaging procedure must give zero for an arising small correction to any unperturbed metric coefficient. The scale
factor is otherwise determined incorrectly. Of course, the backreaction effect due to nonlinearity of general relativity leads to nonzero corrections to the FLRW
background scale factor, but these corrections are of the second order of smallness, so we do not take them into account. Actually, the discussed requirement of
zero average values of first-order cosmological perturbations represents an indispensable condition for the corresponding theory and computer simulations based on
the predicted equations of galaxy dynamics.

However, as we demonstrate explicitly in this paper, there exists a possibility of such matter distributions which lead to nonzero average values of the
first-order metric corrections. Namely, in the framework of the mechanical approach to cosmological problems at the late stage of the Universe evolution we
give a concrete example of a rest mass density profile for which the standard formula determining the scalar perturbations results in their nonzero average
values. Since exactly this formula underlies the modern $N$-body simulations which play an extremely important role for the structure formation analysis, the
discovered weak point must be eliminated in order to be fully confident in their predictions. And we suggest avoiding this challenge without exceeding the
limits of the conventional $\Lambda$CDM model, by cutting off the nonrelativistic gravitational potentials of cosmic bodies/inhomogeneities (e.g., galaxies).

The paper is structured in the following way. First, we enumerate briefly some basic achievements of the mechanical description of cosmological perturbations
which may be associated with discrete cosmology in the nonrelativistic limit. Second, we prove that the commonly used infinite-range gravitational potential can
be characterized by the nonzero average value, and then introduce the finite-range one in order to resolve this problem. Finally, we summarize laconically our
results.

\section{Mechanical description of cosmological perturbations} \setcounter{equation}{0}

According to the mechanical approach, developed in \cite{EZcosm1,EKZ2,EZcosm2} in the framework of the conventional $\Lambda$CDM model (see also the related
recent papers \cite{Eleonora,Clarkson,Chisari,Ruth} where some similar issues and ideas are touched upon in the same spirit), the scalar cosmological
perturbations in the late Universe with flat spatial topology can be described by the perturbed FLRW metric
\be{1} ds^2\approx a^2\left[(1+2\Phi)d\eta^2-(1-2\Phi)\delta_{\alpha\beta}dx^{\alpha}dx^{\beta}\right]\, , \ee
where $a(\eta)$ is the scale factor,
\be{2} \Phi(\eta,{\bf r})=\frac{\varphi({\bf r})}{c^2a(\eta)},\quad \triangle\varphi=4\pi G_N (\rho-\overline\rho)\, , \ee
$\triangle=\delta^{\alpha\beta}\partial^2/(\partial x^{\alpha}\partial x^{\beta})$ stands for the Laplace operator, $G_N$ is the Newtonian gravitational constant,
and $\rho$ represents the rest mass density in the comoving coordinates, being time-independent within the adopted accuracy, while $\overline\rho$ denotes its
constant average value. Here both the nonrelativistic and weak field limits are applied, which means that peculiar velocities of inhomogeneities (galaxies) are
negligibly small in comparison with the speed of light, and the metric corrections are much smaller than the corresponding background metric coefficients (i.e.
$|\Phi|\ll 1$). The function $\Phi$ given by \rf{2} satisfies the following system of linearized Einstein equations of the scalar perturbations theory (see, e.g.,
\cite{Mukhanov,Rubakov}):
\be{3} \triangle\Phi-3{\mathcal H}(\Phi'+{\mathcal H}\Phi)=\frac{4\pi G_N}{c^4}
a^2\left(\delta\varepsilon_{\mathrm{mat}}+\delta\varepsilon_{\mathrm{rad}}\right)\, , \ee
\be{4} \frac{\partial}{\partial x^{\beta}}(\Phi'+{\mathcal H}\Phi)=0\, ,\ee
\be{5} \Phi''+3{\mathcal H}\Phi'+\left(2{\mathcal H}'+{\mathcal H}^2\right)\Phi=\frac{4\pi G_N}{c^4} a^2\delta p_{\mathrm{rad}}\, .\ee

Here the prime denotes the derivative with respect to the conformal time $\eta$, $\mathcal{H}=a'/a$,
$\delta\varepsilon_{\mathrm{rad}}=-3\overline\rho\varphi/a^4$ and $\delta p_{\mathrm{rad}}=\delta\varepsilon_{\mathrm{rad}}/3$ represent the perturbations of
the energy density and pressure of radiation (the corresponding unperturbed/average quantities are neglected) while
$\delta\varepsilon_{\mathrm{mat}}=(\rho-\overline\rho)c^2/a^3+3\overline\rho\varphi/a^4$ is the perturbation of the energy density of the completely
nonrelativistic matter (the corresponding unperturbed/average quantity reads $\overline\varepsilon_{\mathrm{mat}}=\overline\rho c^2/a^3$).

The enumerated results accord with \cite{Landau,Peebles} as well as \cite{gadget2,Vlasov}. In addition to them, let us mention the fact that the equations
expressing the energy conservation (see, e.g., \cite{Rubakov}) hold true with the adopted accuracy for both the nonrelativistic matter and radiation:
\be{6} \delta\varepsilon'_{\mathrm{mat}}+3\mathcal{H}\delta\varepsilon_{\mathrm{mat}}-3\overline\varepsilon_{\mathrm{mat}}\Phi'=0\, ,\ee
\be{7} \delta\varepsilon'_{\mathrm{rad}}+3\mathcal{H}(\delta\varepsilon_{\mathrm{rad}}+\delta p_{\mathrm{rad}})=0\, ,\ee
as one can easily verify by the proper direct substitutions.

In \cite{Peebles} the solution of the Poisson equation \rf{2} for the gravitational potential $\varphi$ (in the comoving coordinates) is presented in the
standard mathematical physics manner as follows:
\be{8} \varphi=-G_N\int_{V'} \frac{\rho({\bf r}')-\overline{\rho}}{|{\bf r}-{\bf r}'|}dV'\, .\ee

Below we focus attention on the extremely important problematic aspect of this commonly used presentation.

\section{Infinite- and finite-range gravitational potentials} \setcounter{equation}{0}

In the case of the infinite-range gravitational potential \rf{8} there is a simple example of the mass distribution leading to nonzero average values of
cosmological perturbations. It bears a direct relation to Einstein-Straus/Swiss-cheese models (see, e.g., the recent papers \cite{ES,SC}). This artificial, but
instructive distribution (described also in \cite{EZcosm1,EZcosm2}) is shown in Fig. 1. The Universe is supposed to be filled with an infinite number of empty
spheres ($\rho=0$) with the exception of point-like masses in their centers, embedded in the homogeneous background ($\rho=\overline\rho$). The radius $R$ of a
given sphere is interconnected with the mass $m$ in its center: $m=4\pi\overline\rho R^3/3$ (all matter from each sphere is concentrated in its center, so the
average rest mass density of such Universe remains equal to $\overline\rho$, as it certainly should be).

\begin{figure}[htbp]
\begin{center}\includegraphics[width=3.22in,height=3.06667in]{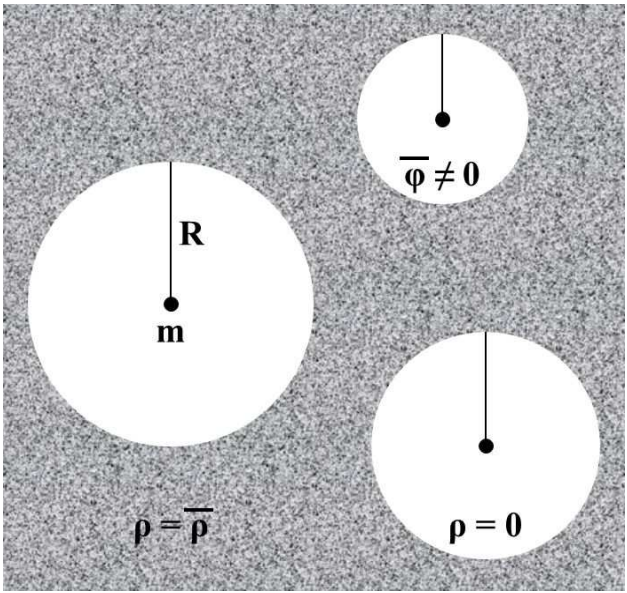}\end{center}
\caption{The mass distribution characterized by the nonzero average value of the infinite-range gravitational potential.}
\end{figure}

In order to determine the gravitational potential $\varphi$ corresponding to the given sphere, one can solve the Poisson equation \rf{2} with the appropriate
boundary conditions $\varphi(R)=0$, $d\varphi/dr(R)=0$ or use the standard prescription \rf{8}. The result is the same: inside the sphere (the region I)
\be{9} \varphi_{\mathrm{I}}=2\pi G_N\overline{\rho}\left(R^2-\frac{r^2}{3}\right)-\frac{G_Nm}{r}\, , \ee
while outside the sphere (the region II)
\be{10} \varphi_{\mathrm{II}}=\frac{4}{3}\pi \overline{\rho}R^3\frac{G_N}{r}-\frac{G_Nm}{r}=0\, .\ee

Averaging the derived function \rf{9} over the volume $V=4\pi R^3/3$ of the sphere, we immediately obtain the senseless result
\be{11} \overline\varphi_{\mathrm{I}}=\frac{1}{V}\int_V \varphi_{\mathrm{I}} dV=-\frac{3G_Nm}{10R}\neq0\ee
meaning that the standard prescription \rf{8} can lead to unreasonable nonzero average values of cosmological perturbations. One can naively suppose that the
result \rf{11} is true only for the considered region of the finite volume $V$, while averaging over the infinite volume saves the situation. This
argumentation is apparently wrong since there is an infinite number of such regions in the model under consideration, and each of them makes a nonzero
(negative) contribution when averaging over the infinite volume. Thus, the average value will be again nonzero (negative).

Trying to save the situation in a different way, one can also change the boundary condition $\varphi(R)=0$ for the Poisson equation \rf{2}. Namely, one can
require that $\varphi(R)=\mathrm{const}\neq0$ instead, and then adjust this constant in order to satisfy the desired condition $\overline\varphi=0$ when
averaging over the infinite volume containing an infinite number of regions depicted in Fig.~1. Actually, such a change is equivalent to simply adding this
constant in the right-hand side of \rf{8}. Again, these arguments are apparently wrong. The zero boundary value $\varphi(R)=0$ is in concordance with the
standard formula \rf{8}, which, in its turn, agrees with the generally accepted mathematical physics requirement that any fluctuation $\delta\rho$ produces a
decreasing gravitational potential vanishing at spatial infinity. In this connection, if one changes the boundary value or, equivalently, adds some nonzero
constant in the right-hand side of \rf{8}, then such an additional term has unclear physical interpretation and no evident source. It breaks the superposition
principle and certainly contradicts a natural demand of vanishing cosmological perturbations in the absence of inhomogeneities (i.e. when $\delta\rho=0$).

Of course, since the function $\Phi$ describes the deviation of the metric coefficients in \rf{1} from the corresponding average quantities, its own average
value must be equal to zero: $\overline{\Phi}=0$. The same statement must hold true for $\varphi$, $\delta\varepsilon_{\mathrm{rad}}$, etc. The discovered
indubitable disadvantage of the formula \rf{8} should not be ignored in the modern $N$-body simulations (along with \cite{Peebles,gadget2}, see \cite{review}).

It is important to remark that the nonzero time-dependent contribution $\overline{\Phi}$ in the averaged metric
$$
a^2\left[(1+2\overline\Phi)d\eta^2-(1-2\overline\Phi)\delta_{\alpha\beta}dx^{\alpha}dx^{\beta}\right]
$$
cannot be eliminated by a coordinate transformation $\eta\rightarrow\eta+\epsilon(\eta)$, $x^{\alpha}\rightarrow x^{\alpha}(1-\lambda)$, where $\epsilon(\eta)$ is
some function of the first order of smallness and $\lambda$ is some constant of the same order \cite{transform1,transform2,transform3}. Really, if such a
possibility of elimination had existed, then, according to \cite{transform2}, the following equations would have been true:
$\overline\Phi=\epsilon'+\mathcal{H}\epsilon$ and $\overline\Phi=\lambda-\mathcal{H}\epsilon$. However, taking into account that, as it follows directly from
\rf{2}, $\overline\Phi\sim1/a$, one can easily prove that these equations are consistent only in the case $\mathcal{H}'=\mathcal{H}^2$, and this equality does not
hold true in the Universe, which is supposed to be filled not only with the dark energy in the form of the $\Lambda$-term, but also with the nonrelativistic
matter. The latter, as is generally known, makes different contributions in the background Friedmann equations for $\mathcal{H}^2$ and $\mathcal{H}'$. Thus, the
considered coordinate transformation does not help as well.

In order to avoid this difficulty, let us introduce the finite-range gravitational potential by modifying \rf{8} as follows:
\be{12} \varphi=-G_N\int_{V'} \frac{\rho({\bf r}')-\overline{\rho}}{|{\bf r}-{\bf r}'|}\Theta(R_*-|{\bf r}-{\bf r}'|)dV'\, ,\ee
where $\Theta$ represents the Heaviside step function, and $R_*$ is some positive (generally speaking, time-dependent) cutoff distance which may be associated
with the particle horizon. This modification is inspired by a similar cutoff when describing propagation of electro-magnetic or gravitational waves: the field is
nonzero only in those points which have received the corresponding signal, even if its source is resting. Then the taken step may be interpreted as making the
Newtonian classical mechanics more precise by supplementing it with the special (not general!) relativity idea of the signal propagation speed finiteness. Here
peculiar velocities of inhomogeneities (galaxies) are completely neglected as before, and the first-order perturbation theory holds true (the gravitational field
described by the metric corrections remains weak). Evidently, the introduction of the formula \rf{12} instead of the predecessor without the Heaviside step
function represents a particular modification of the gravitation theory where scalar modes are no longer instantaneous.

Applying \rf{12} to the mass distribution under consideration and restricting ourselves to the case $R_*>2R$, we get \rf{9} and \rf{10} for the gravitational
potentials $\varphi_{\mathrm{I}}$ and $\varphi_{\mathrm{II}}$ in the regions I ($0<r<R$) and II ($R<r<R_*-R$) respectively, as well as
\be{13} \varphi_{\mathrm{III}}=\frac{\pi G_N\overline{\rho}}{r}\left[R^2\left(R_*-r\right)-\frac{2}{3}R^3-\frac{1}{3}\left(R_*-r\right)^3\right]\ee
in the region III ($R_*-R<r<R_*$), while
\be{14} \varphi_{\mathrm{IV}}=\frac{\pi G_N\overline{\rho}}{r}\left[R^2\left(R_*-r\right)+\frac{2}{3}R^3-\frac{1}{3}\left(R_*-r\right)^3\right]\ee
in the region IV ($R_*<r<R_*+R$). Finally,
\be{15} \varphi_{\mathrm{V}}=0\ee
in the region V ($r>R_*+R$).

Now the direct calculation gives the desired result:
\ba{16} \int\limits_0^{R}\varphi_{\mathrm{I}}(r)r^2dr&+&\int\limits_R^{R_*-R}\varphi_{\mathrm{II}}(r)r^2dr\nn\\
+\int\limits_{R_*-R}^{R_*}\varphi_{\mathrm{III}}(r)r^2dr&+&\int\limits_{R_*}^{R_*+R}\varphi_{\mathrm{IV}}(r)r^2dr=0\, ,\ea
therefore, $\overline\varphi=0$, as it certainly should be. Thus, the use of the finite-range gravitational potential \rf{12} instead of the infinite-range one
\rf{8} leads to reasonable zero average values of cosmological perturbations. This advantage of the proposed formula \rf{12} in comparison with \rf{8} may be
taken into account when simulating the behavior of $N$-body systems.

Let us mention that in the framework of the extension of the $\Lambda$CDM model, assuming the presence in the Universe of the additional constituent (namely,
quintessence) with the linear equation of state $\varepsilon_q=\omega_qp_q$ with the constant parameter $\omega_q=-1/3$, the discussed problem of nonzero
average values of cosmological perturbations in the case of the infinite-range gravitational potential is resolved in a different manner: quintessence
fluctuations around a point-like nonrelativistic matter inhomogeneity cause the Yukawa form of its potential instead of the Newtonian one, and the average
value of the total potential produced by all inhomogeneities is really zero \cite{BEZ1}, irrespective of the interaction range and its cutoff.

The other extension, assuming the negative spatial curvature, is also characterized by the potential of a point-like inhomogeneity, similar to the Yukawa one,
so the average value of the total potential is again zero \cite{EZcosm1}.

Returning to the conventional model under consideration and introducing the dimensionless quantities $\chi=r/R$, $\xi=R_*/R$ and
$\widetilde{\varphi}=\varphi/\left(4 \pi G_N\overline\rho R^2/3\right)$, from \rf{9}, \rf{13} and \rf{14} we obtain respectively
\be{17} \widetilde{\varphi}_{\mathrm{I}}(\chi)=\frac{3}{2}-\frac{\chi^2}{2}-\frac{1}{\chi}\, ,\ee
\be{18} \widetilde{\varphi}_{\mathrm{III}}(\chi)=\frac{3(\xi-\chi)}{4\chi}-\frac{1}{2\chi}-\frac{\left(\xi-\chi\right)^3}{4\chi}\, ,\ee
\be{19} \quad \widetilde{\varphi}_{\mathrm{IV}}(\chi)=\frac{3(\xi-\chi)}{4\chi}+\frac{1}{2\chi}-\frac{\left(\xi-\chi\right)^3}{4\chi}\, ,\ee
while from \rf{10} and \rf{15} it trivially follows that $\widetilde{\varphi}_{\mathrm{II}}=\widetilde{\varphi}_{\mathrm{V}}=0$. The spherical surfaces $r=R;\,
R_*-R;\, R_*+R$ (where the function $\varphi(r)$ is smooth, as one can easily demonstrate) correspond to $\chi=1;\, \xi-1;\, \xi+1$ (where the smoothness
conditions are satisfied for the function $\widetilde\varphi(\chi)$). At the same time on the surface $r=R_*$ corresponding to $\chi=\xi$ and delimiting the
spatial regions III and IV there is a foreseeable jump discontinuity:
\be{20} \widetilde{\varphi}_{\mathrm{III}}(\xi)-\widetilde{\varphi}_{\mathrm{IV}}(\xi)=-\frac{1}{\xi},\quad
\frac{d\widetilde{\varphi}_{\mathrm{III}}}{d\chi}(\xi)-\frac{d\widetilde{\varphi}_{\mathrm{IV}}}{d\chi}(\xi)=\frac{1}{\xi^2}\, .\ee

Really, on the surface under consideration the action of the central mass gravitational field ends, so the result \rf{20} is expected, and it does not relate
to the modification \rf{12} itself. In terms of the function $\varphi(r)$ and its derivative $d\varphi/dr$ this result can be trivially rewritten as follows:
\ba{21} &{}&{\varphi}_{\mathrm{III}}(R_*)-{\varphi}_{\mathrm{IV}}(R_*)=-\frac{G_Nm}{R_*}\, ,\nn\\
&{}&\frac{d{\varphi}_{\mathrm{III}}}{dr}(R_*)-\frac{d{\varphi}_{\mathrm{IV}}}{dr}(R_*)=\frac{G_Nm}{R_*^2}\, .\ea

The dependence $\widetilde\varphi(\chi)$ is depicted in Fig. 2 (for solely illustrative purposes the choice $\xi=5$ is made).

\begin{figure}[htbp]
\begin{center}\includegraphics[width=3.3in,height=2.1in]{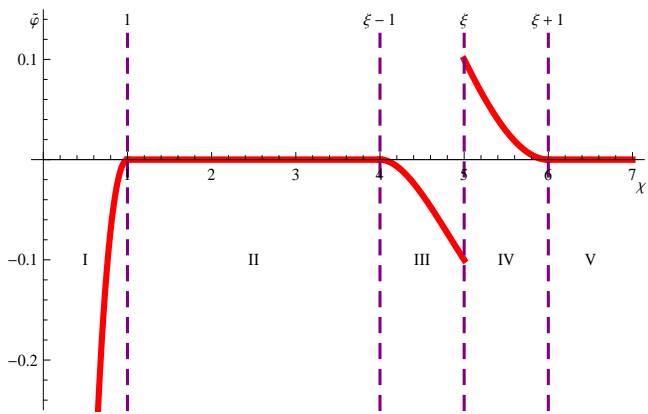}\end{center}
\caption{The finite-range gravitational potential as a function of the radial distance.}
\end{figure}

\section{Conclusion}

We have proven that for some matter distributions at the late stage of the Universe evolution (see, e.g., Fig. 1) the nonrelativistic gravitational potential can
be characterized by the nonzero average value if the standard formula \rf{8} is applied. This situation is absolutely inadmissible, so in order to resolve this
challenge we have cut off the gravitational potential and with the help of the modified formula \rf{12} obtained the desired result $\overline\varphi=0$.
Evidently, it is valid not only for the considered concrete density profile but also for an arbitrary one. It is important to stress that our conclusion is
correct provided that the made assumptions concerning the FLRW geometry as an average one and the comoving volume averaging as an appropriate averaging procedure
are valid. Then the suggested application of the finite-range potentials instead of the infinite-range ones can improve the quality and precision of the
cosmological simulations with respect to their representation of the physical reality.

\section*{Acknowledgements}

The work of M. Eingorn and B.~Vlahovic was supported by NSF CREST award HRD-1345219 and NASA grant NNX09AV07A.

We would like to thank the Referee for critical remarks which have considerably improved the discussion parts of the manuscript.

M. Eingorn and B. Vlahovic are also grateful to V.~Gurzadyan and S.~Matinyan for useful discussions and valuable comments underlying the given investigation.


\begin{thebibliography}{}
%
\bibitem{EZcosm2}
M. Eingorn and A. Zhuk, {\em Remarks on mechanical approach to observable Universe}, JCAP {\bf 05} (2014) 024; [astro-ph/1309.4924].
%
%
\bibitem{EZcosm1}
M. Eingorn and A. Zhuk, {\em Hubble flows and gravitational potentials in observable Universe}, JCAP {\bf 09} (2012) 026; [astro-ph/1205.2384].
%
%
\bibitem{EKZ2}
M. Eingorn, A. Kudinova and A. Zhuk, {\em Dynamics of astrophysical objects against the cosmological background}, JCAP {\bf 04} (2013) 010;
[astro-ph/1211.4045].
%
%
\bibitem{Eleonora}
E. Villa, S. Matarrese and D. Maino, {\em Cosmological dynamics: from the Eulerian to the Lagrangian frame. Part I. Newtonian approximation}, JCAP {\bf 06}
(2014) 041; [astro-ph/1403.6806].
%
%
\bibitem{Clarkson}
O. Umeh, C. Clarkson and R. Maartens, {\em Nonlinear relativistic corrections to cosmological distances, redshift and gravitational lensing magnification. II -
Derivation}, Class. Quant. Grav. {\bf 31} (2014) 205001; [astro-ph/1402.1933].
%
%
\bibitem{Chisari}
N.E. Chisari and M. Zaldarriaga, {\em Connection between Newtonian simulations and general relativity}, Phys. Rev. D {\bf 83} (2011) 123505;
[astro-ph/1101.3555].
%
%
\bibitem{Ruth}
J. Adamek, D. Daverio, R. Durrer and M. Kunz, {\em General Relativistic N-body simulations in the weak field limit}, Phys. Rev. D {\bf 88} (2013) 103527;
[astro-ph/1308.6524].
%
%
\bibitem{Mukhanov}
V.F. Mukhanov, H.A. Feldman and R.H. Brandenberger, {\em Theory of cosmological perturbations}, Physics Reports {\bf 215} (1992) 203.
%
%
\bibitem{Rubakov}
D.S. Gorbunov and V.A. Rubakov, {\em Introduction to the Theory of the Early Universe: Cosmological Perturbations and Inflationary Theory}, World Scientific,
Singapore (2011).
%
%
\bibitem{Landau}
L.D. Landau and E.M. Lifshitz, {\em The Classical Theory of Fields (Course of Theoretical Physics Series, V. 2)}, Oxford Pergamon Press, Oxford (2000).
%
%
\bibitem{Peebles}
P.J.E. Peebles, {\em The large-scale structure of the Universe}, Princeton University Press, Princeton (1980).
%
%
\bibitem{gadget2}
V. Springel, {\em The cosmological simulation code GADGET-2}, MNRAS {\bf 364} (2005) 1105; [astro-ph/0505010].
%
%
\bibitem{Vlasov}
V. Rubakov and A. Vlasov, {\em What do we learn from CMB observations}, Physics of Atomic Nuclei {\bf 75} (2012) 1123; [astro-ph/1008.1704].
%
%
\bibitem{ES}
M. Mars, F.C. Mena and R. Vera, {\em Review on exact and perturbative deformations of the Einstein-Straus model: uniqueness and rigidity results}, Gen. Rel. Grav.
{\bf 45} (2013) 2143; [gr-qc/1307.4371].
%
%
\bibitem{SC}
P. Fleury, {\em Swiss-cheese models and the Dyer-Roeder approximation}, JCAP {\bf 06} (2014) 054; [astro-ph/1402.3123].
%
%
\bibitem{review}
F. Bernardeau, S. Colombi, E. Gaztanaga and R. Scoccimarro, {\em Large-scale structure of the Universe and cosmological perturbation theory}, Physics Reports
{\bf 367} (2002) 1.
%


\bibitem{transform1}
S. Weinberg, {\em Adiabatic modes in cosmology}, Phys. Rev. D {\bf 67} (2003) 123504; [astro-ph/0302326].

\bibitem{transform2}
P. Creminelli, C. Pitrou and F. Vernizzi, {\em The CMB bispectrum in the squeezed limit}, JCAP {\bf 11} (2011) 025; [astro-ph/1109.1822].

\bibitem{transform3}
M. Mirbabayi and M. Zaldarriaga, {\em CMB anisotropies from a gradient mode}, JCAP {\bf 03} (2015) 056; [astro-ph/1409.4777].


%
\bibitem{BEZ1}
A. Burgazli, M. Eingorn and A. Zhuk, {\em Rigorous theoretical constraint on constant negative EoS parameter $\omega$ and its effect for the late Universe},
EPJC {\bf 75} (2015) 118; [astro-ph/1301.0418].
%
%
%
%
%
\end{thebibliography}
\end{document}